# Fusion of Machine Learning and Blockchain-based Privacy-Preserving Approach for Health Care Data in the Internet of Things

**Behnam Rezaei Bezanjani**[1] . **Seyyed Hamid Ghafouri**[1*] . **Reza Gholamrezaei**[1]

**Abstract:** In recent years, the rapid integration of Internet of Things (IoT) devices into the healthcare sector has brought about revolutionary advancements in patient care and data management. While these technological innovations hold immense promise, they concurrently raise critical security concerns, particularly in safeguarding medical data against potential cyber threats. The sensitive nature of health-related information requires robust measures to ensure patient data's confidentiality, integrity, and availability within IoT-enabled medical environments. Addressing the imperative need for enhanced security in IoT-based healthcare systems, we propose a comprehensive method encompassing three distinct phases. In the first phase, we implement Blockchain-Enabled Request and Transaction Encryption to fortify the security of data transactions, providing an immutable and transparent framework. Subsequently, in the second phase, we introduce Request Pattern Recognition Check, leveraging diverse data sources to identify and thwart potential unauthorized access attempts. Finally, the third phase incorporates Feature Selection and the BiLSTM network to enhance the accuracy and efficiency of intrusion detection through advanced machine-learning techniques. We compared the simulation results of the proposed method with three recent related methods namely AIBPSF-IoMT, OMLIDS-PBIoT, and AIMMFIDS. The evaluation criteria encompass detection rates, false alarm rates, precision, recall, and accuracy, crucial benchmarks in assessing the overall performance of intrusion detection systems. Notably, our findings reveal that the proposed method outperforms these existing methods across all evaluated criteria, underscoring its superiority in enhancing the security posture of IoT-based healthcare systems.

---

✉ Behnam Rezaei Bezanjani
B.rezai@kmu.ac.ir

✉ Seyyed Hamid Ghafouri*
Ghafoori@iauk.ac.ir

✉ Reza Gholamrezaei
Gholamrezaei@iauk.ac.ir

[1]Department of Computer Engineering, Kerman Branch, Islamic Azad University, Kerman, Iran.

# 1 Introduction

The proliferation of Internet of Things (IoT) technologies has revolutionized various industries, offering unprecedented connectivity and convenience. However, the widespread adoption of IoT systems has also raised concerns about security vulnerabilities and potential cyberattacks [4]. To address these challenges, researchers have explored innovative solutions at the intersection of blockchain (BC) technology and deep learning (DL), aiming to enhance the security and efficiency of IoT ecosystems. BC technology, originally devised as the underlying framework for cryptocurrencies, has emerged as a robust and decentralized approach to secure data transactions [2]. Its application in the realm of IoT introduces a new paradigm where trust, transparency, and immutability play pivotal roles in safeguarding sensitive information. Recent studies, such as GTxChain [1] and AIBPSF-IoMT [10], showcase the development of secure, intelligent BC IoT architectures that leverage advanced techniques, including graph neural networks and artificial intelligence, to fortify the integrity of IoT networks. On the other hand, DL has garnered significant attention across various domains for its unparalleled accuracy in data analysis and pattern recognition [6,7]. With applications spanning healthcare, medical image processing, and cybersecurity [8,9], DL algorithms have demonstrated their prowess in addressing complex challenges. Notably, the integration of BC and DL has been explored in recent research, aiming to harness both technologies' strengths for enhanced security and predictive capabilities [11,12]. Figure 1 shows machine learning and BC in the Internet of Medical Things environment (IoMT).

This paper introduces an innovative approach to address the critical challenge of medical data security in the IoT landscape. Our method seamlessly integrates BC technology and advanced machine-learning techniques. Specifically, BiLSTM (Bidirectional Long Short-Term Memory) provides a robust framework for enhancing the confidentiality and integrity of sensitive healthcare information. We proposed a comprehensive approach with three distinct phases to address the urgent need to improve security in IoT-based healthcare systems. In the first phase, we evaluated the reliability of IoT devices through reputation-based trust estimation and off-chain data storage. The second phase uses blockchain technology to prevent attacks and authenticate data. In the third phase, artificial intelligence, specifically BiLSTM, is used to classify attacks.

**Unique Aspects of the Integration:**

1. Integrating BiLSTM with Blockchain is a novel approach that leverages the strengths of both technologies. While Blockchain ensures the security and immutability of data, BiLSTM excels at handling sequential data, making it highly effective for detecting patterns and anomalies in healthcare data.
2. BiLSTM's ability to process data in both forward and backward directions improves the accuracy and reliability of attack detection, outperforming traditional machine learning methods that do not consider the temporal dependencies in the data.
3. Blockchain technology adds a layer of security by decentralizing data storage and providing a transparent and tamper-proof ledger, significantly enhancing data integrity and trust.

To enhance the scalability and efficiency of the proposed system, Cloud, Edge, and Fog computing paradigms are incorporated as shown in the Figure 1.

**Cloud Computing**: The cloud layer is responsible for intensive data processing and long-term storage. It manages large-scale analytics and supports global insights derived from aggregated data.

**Edge Computing**: The edge layer brings computation closer to the data sources. It handles time-sensitive tasks, reduces latency, and decreases the bandwidth required for data transmission to the cloud. Devices and gateways in this layer perform initial data filtering and preliminary analysis.

**Fog Computing**: Positioned between the edge and the cloud, the fog layer supports distributed computing resources and storage. It enables intermediate data processing, alleviating the computational load on both edge devices and cloud servers. Fog nodes ensure real-time processing capabilities and enhance the system's responsiveness.

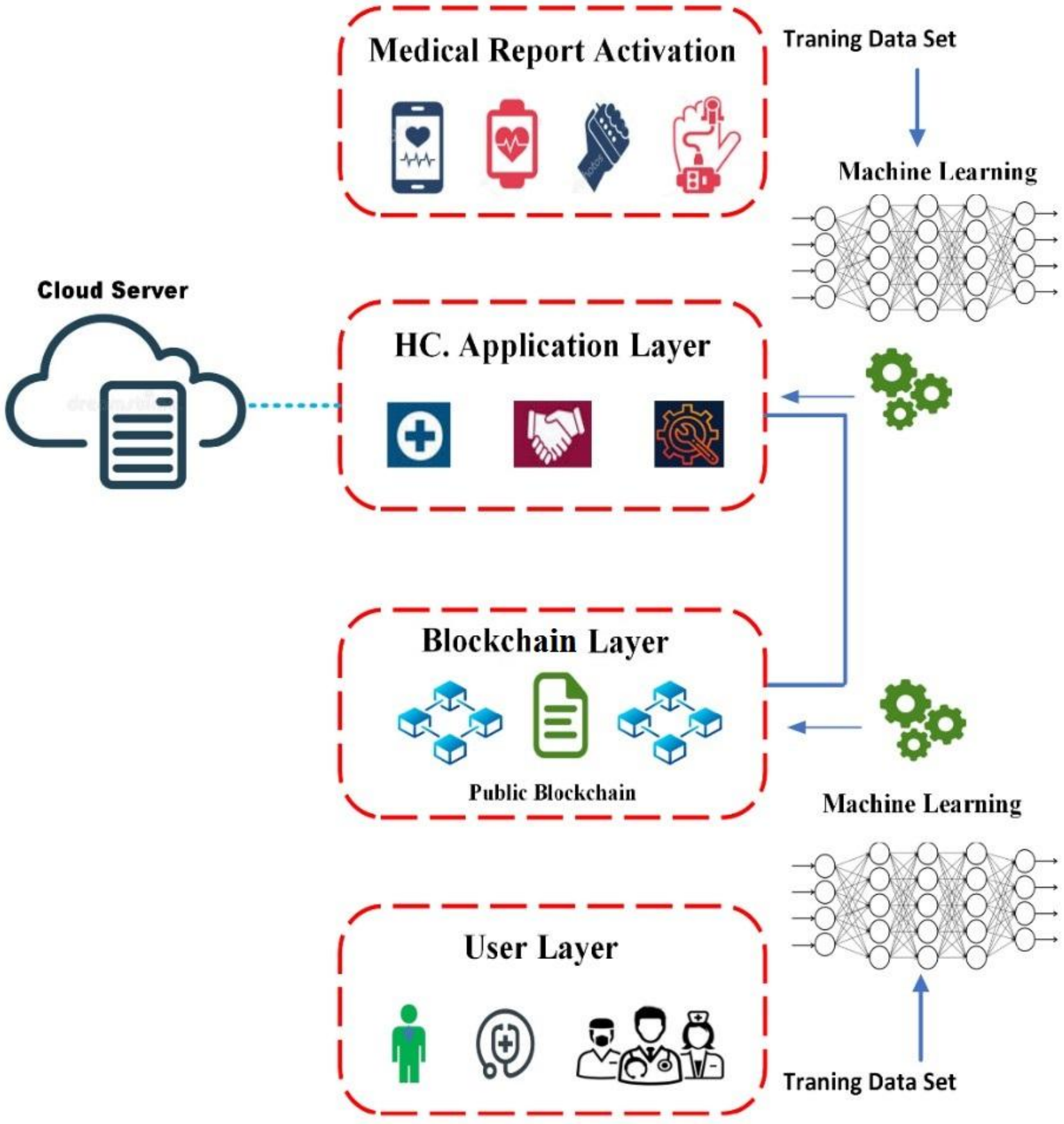

**Fig. 1** ML and BC in the IoMT.

Despite the solutions provided by researchers, there are still problems that need to be addressed and solved. We have stated these problems in Section 1-1.

## 1.1 Problem statement

Despite significant advancements in the field, several critical issues remain unaddressed:

1. **Inability to Predict New Attacks**: Conventional ML techniques, while accurate, often fail to detect novel and complex threats due to their limited adaptability to new and sophisticated attack patterns [1-4].

2. **Lack of Generalizability**: Previous research has typically focused on a narrow spectrum of attack types, which hinders the comprehensive detection of diverse attack patterns and leaves significant security gaps [10,20,22].

## 1.1 Contributions

Our main contributions are unfolded as follows:

1. **Enhancing Medical Data Security**: We propose an innovative method that integrates BC technology with a BiLSTM network to enhance data privacy and security, specifically designed to tackle the issue of detecting novel and complex threats.
2. **Preventing Overfitting and Reducing Training Time**: Our method employs a feature selection approach that prevents overfitting and reduces the training and testing time of the BiLSTM network while maintaining high accuracy. This directly addresses the challenge of limited generalizability by improving the model's efficiency and adaptability.
3. **Predicting Multiple Attack Types**: The proposed method has been rigorously evaluated on two different datasets and shows superior performance in predicting more than ten different types of attacks, thus ensuring a broader and more comprehensive coverage of potential threats.

Table 1 summarizes the comparison of previous methods with the proposed method.

**Table 1.** Comparison of previous methods with the proposed method.

| Method | Precision | Recall | F1-score | Accuracy | Detection rate | False alarm rate |
|---|---|---|---|---|---|---|
| [10] | ✓ | × | × | ✓ | ✓ | × |
| [33] | ✓ | × | × | ✓ | ✓ | × |
| [34] | × | ✓ | × | ✓ | × | ✓ |
| [13] | ✓ | × | × | ✓ | ✓ | × |
| [16] | × | ✓ | × | ✓ | × | ✓ |
| **Proposed method** | ✓ | ✓ | ✓ | ✓ | ✓ | ✓ |

Some abbreviations are used in the text. Table 2 shows the description of these abbreviations.

**Table 2.** Detailed Description of Abbreviations Used Throughout the Study.

| **Symbol** | **Description** | **Symbol** | **Description** |
|---|---|---|---|
| IoT | Internet of Things | RNN | Recurrent Neural Network |
| BiLSTM | Bidirectional Long Short-Term Memory | TP | True Positive |
| IoV | Internet of Vehicles | TN | True Negative |
| ODDS | Optimized Dynamic Data Storage | FP | False Positive |
| WOA | Whale Optimization Algorithm | FN | False Negative |
| ML | Machine Learning | DL | Deep Learning |
| BC | Blockchain | IoMT | Internet of Medical Things |

## 1.2 Paper Organization

The subsequent sections will be structured as follows: Section II will assess and analyze previous studies and relevant research. Section III outlines the proposed method, including its steps and specific details. Section IV presents a thorough analysis of the simulation results. Finally, Section V gives the future works and conclusion of the research paper.

# 2 Related work

Several methods using BC and DL have been proposed in recent years to maintain data security and privacy. In this section, we review and compare related works.

An advanced approach to detect botnet attacks in industrial IoT systems is proposed in [13]. Their method uses multi-layer DL and represents a proactive strategy to mitigate security threats in wireless communication and mobile computing environments. This study focuses on strengthening the robustness of industrial IoT systems against evolving cyber threats, especially threats related to botnet attacks. In [14], the authors introduced a new method called BlockMedCare, a new healthcare system that integrates IoT, BC, and InterPlanetary File System (IPFS). Their work addresses data management security concerns in healthcare using BC technology. The system aims to ensure the integrity and confidentiality of healthcare data and contributes to the broader discussion of increasing security in healthcare informatics. An energy-efficient distributed light encryption and authentication method is presented in [15], designed to enhance security in IoT communications. Focusing on energy-constrained IoT devices, this approach contributes to developing secure and resource-efficient distributed systems. This study emphasizes the importance of energy-aware security measures in the context of the Internet of Things.

In [16], an anomaly detection method based on an iterative convolution autoencoder is proposed for IoT time series data. This approach combines DL and sequential modeling to identify unusual patterns in IoT data streams. This work addresses the critical need for robust anomaly detection mechanisms in IoT applications and ensures the reliability of time series data analysis. Another research [17] introduced a blockchain-based authentication and key agreement protocol for the Internet of Vehicles (IoV). Their method uses roadside units to enhance communication security in IoV. This study uses BC technology to provide a secure and decentralized framework for authentication and key agreement in automotive communication systems. In [18], a secure fusion approach for IoT in intelligent autonomous multi-robot systems is proposed. Focusing on security in multi-robot systems, their approach ensures secure data integration in IoT-enabled robotic environments. This work contributes to the field of autonomous systems by addressing the security challenges associated with data integration in a multi-robot context.

Blockchain-based optimized dynamic data storage (ODSD) in IoT for wireless sensor networks is presented in [19]. The optimization strategy aims to increase data storage efficiency in Internet of Things applications, especially in wireless sensor networks. This work addresses data storage and management challenges and helps improve overall performance in IoT scenarios. In [20], the application of BC technology for secure communication of healthcare data in 5G IoT networks is reviewed. This study focuses on ensuring the secure exchange of healthcare data among non-terminal nodes in 5G networks. This will help increase the security and privacy of healthcare data in advanced communication infrastructures. A distributed network security framework for the Internet of Energy based on the Internet of Things is proposed in [21]. This work addresses security challenges in the energy sector and provides a comprehensive framework for enhancing Internet of Energy security through IoT-based security measures. This study emphasizes the importance of protecting critical infrastructure in the age of the Internet of Things.

In [22], an overview of smart contracts and challenges, developments, and platforms are discussed. A comprehensive overview helps to understand smart contract technology, addresses key challenges, and showcases advances. This work provides valuable insights for researchers and practitioners interested in developing and deploying smart contracts. In another research [23], a blockchain-based, decentralized, federated learning framework [38] with committee consensus is presented. Their approach focuses on addressing privacy and security concerns in federated learning scenarios. This study contributes to developing decentralized and secure federated learning frameworks by applying BC and committee consensus. In [24], a Secure Evidence and Incident Management Framework (SIEMF) for the Internet of Vehicles is introduced. This framework combines DL and BC technologies to manage incidents and evidence in connected vehicles. This work aims to improve the security and forensic capabilities in a car environment equipped with the Internet of Things.

In another research [25], the authors have proposed mixed localization-based outlier models to secure data migration in cloud centers. This work focuses on increasing the security of data transfer processes in cloud environments. Using a hybrid localization approach, this study addresses outliers and contributes to the overall security of data migration in cloud centers. In [26], utilizing a distributed blockchain-based approach, a protection framework for modern power systems against cyber-attacks is introduced. This work revolves around strengthening the cyber security of modern power systems using BC technology. This framework aims to protect power systems from cyber threats through distributed and secure data protection measures. The authors present an elastic and cost-effective data carrier architecture for smart contracts on the BC [27]. Focusing on smart contract applications, this study addresses scalability and cost concerns in blockchain-based systems. The proposed architecture increases the efficiency and cost-effectiveness of data carriers in implementing smart contracts.

In [28], a distributed attack detection framework based on semi-supervised learning is proposed for the Internet of Things. This framework aims to improve attack detection in IoT environments using semi-supervised learning techniques. This approach increases the robustness of attack detection in distributed IoT scenarios by using labeled and unlabeled data. An intrusion detection system based on DL is introduced in [29]. Focusing on network security, it uses DL techniques to enhance intrusion detection. This study contributes to the field of intrusion detection by addressing the challenges posed by adversaries and enhancing DL capabilities in security applications.

**Table 3.** Comparative Analysis of Different Approaches Highlighting Their Advantages and Disadvantages.

| Ref. | Year | Advantages | Disadvantages |
|---|---|---|---|
| [12] | 2023 | Detailed taxonomy and identification of challenges; useful for guiding future research. | Lacks specific solutions or case studies addressing the challenges identified |

| Ref. | Year | Advantages | Disadvantages |
|---|---|---|---|
| [13] | 2022 | High accuracy and low false positive rates in botnet attack detection. | Focused on a specific type of attack, which may limit generalizability to other security threats. |
| [14] | 2022 | Improved data security and management efficiency in healthcare applications. | Implementation complexity and potential scalability issues in large-scale systems. |
| [16] | 2020 | High detection accuracy for anomalies in IoT time series data. | It is computationally intensive, which may limit real-time application. |
| [17] | 2021 | Enhanced security and efficiency in vehicular communications. | Potentially high implementation costs and complexity. |
| [18] | 2021 | Improved data security and system performance in multi-robot systems. | Limited to smart autonomous systems; scalability concerns. |
| [19] | 2021 | Enhanced data retrieval speed and storage optimization in IoT. | Specific to wireless sensor networks, limiting broader applicability. |
| [21] | 2021 | Improved network security and system robustness for the energy internet. | Potentially high implementation costs and integration challenges. |
| [25] | 2019 | Enhanced data integrity and security during cloud-based data migration. | Potentially high computational requirements for large datasets. |
| [26] | 2018 | Improved performance and reduced costs in smart contract execution. | Focused on power systems, which may limit applicability to other domains. |
| [29] | 2018 | Significant improvements in detection accuracy and system resilience against adversarial attacks. | Adversarial training can be computationally intensive. |

## 3 Proposed Method

In an era where technological advancements are reshaping the healthcare landscape, the integration of IoT in the medical domain has ushered in unprecedented possibilities for data-driven insights and patient care. However, with this proliferation of connected devices and the seamless exchange of medical data, the paramount concern becomes ensuring the robust security and privacy of sensitive information. This paper delves into a comprehensive approach to safeguarding medical data within the IoT framework. By exploring the synergy of BC technology, the modified binary whale method for feature selection, and the BiLSTM network, we aim to elucidate an integrated strategy that not only fortifies the confidentiality and integrity of medical data but also ensures the resilience of the overall healthcare ecosystem in the face of evolving cyber threats. The framework of the proposed method (Figure 2) includes two components. The interoperability of the Blockchain-Enabled AI Security Agent is illustrated in a three-phase process:

1. **Phase 1**: **Blockchain-Enabled Request and Transaction Encryption**: This phase ensures that all requests and transactions are cryptographically signed and authenticated using the blockchain platform. It provides an immutable and transparent framework for secure data transactions.

2. **Phase 2: Request Pattern Recognition Check**: In this phase, the data source verifies the pattern of incoming requests. It checks if the request pattern is recognized and provides the relevant security details. If the pattern is not recognized, the system informs the agent and provides details of similar patterns. This helps in early detection and prevention of unauthorized access.
3. **Phase 3: ML/DL Algorithm**: Features are selected using the modified WOA before training the model. This step is crucial to enhancing the intrusion detection system's accuracy and efficiency. The selected features are then processed by the BiLSTM network, which decides on the legitimacy of the requests.

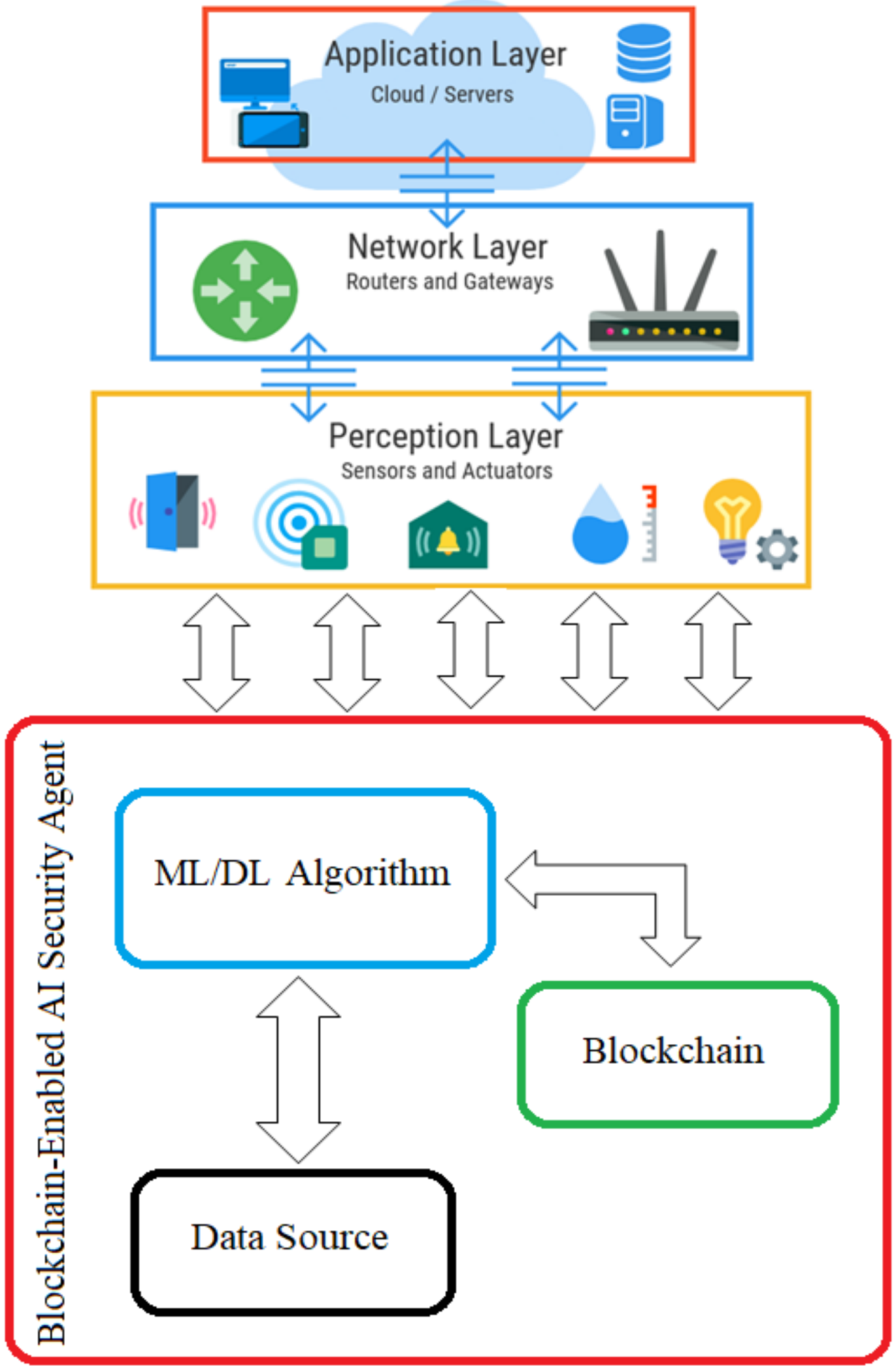

**Fig. 2** Framework of the proposed method.

| Algorithm 1. Pseudocode for the proposed AI and Blockchain-based Predictive Security Framework |
| --- |
| 1. Initialize Blockchain:<br>a. Set up blockchain network<br>b. Define smart contracts for data integrity and security<br><br>2. Data Collection:<br>a. Collect data from IoMT devices<br>b. Preprocess data (cleaning, normalization, etc.) |

3. Feature Extraction:
   a. Extract relevant features from preprocessed data
   b. Perform feature selection and dimensionality reduction

4. Train Machine Learning Model:
   a. Split data into training and testing sets
   b. Select appropriate ML algorithm (e.g., SVM, Random Forest)
   c. Train the model on training data
   d. Validate the model using cross-validation techniques

5. Deploy Model on Edge Devices:
   a. Deploy trained model on edge devices for real-time processing
   b. Implement local inference for anomaly detection

6. Blockchain Integration:
   a. Record detected anomalies and device data on blockchain
   b. Use smart contracts to trigger alerts and automated responses.

## 3.1 Blockchain-Enabled AI Security Agent

In this section, we will explore the collaborative workings of the agent with IoT devices. The agent's functionality and operations are designed to establish a seamless partnership with IoT devices, ensuring secure and efficient communication. The agent operates through a collaborative effort of multiple components, working in unison to establish secure communication channels with IoT devices. Once an external device sends a request, the agent springs into action, diligently scrutinizing the legitimacy of each incoming request to safeguard the IoT device's underlying platform. Comprised of three essential components - a BC platform, a data source, and an ML/DL algorithm- the agent orchestrates a series of efficient operations to ensure the seamless and secure flow of data to the IoT device. Next, we explain each of the components. The flowchart of how the agent works is shown in Figure 3.

## 3.2 Phase 1. Blockchain-Enabled Request and Transaction Encryption

Acting as the agent's initial component, the BC platform assumes a vital role in guaranteeing the cryptographically signed nature of all transactions and requests. Its primary aim is to authenticate requests originating from external sources beyond the IoT architecture. When an outsider sends a request, the device communicates with the BC to verify the integrity of the request and ensure that it remains unchanged. After this verification process, a signed message is generated. This information is then forwarded to the subsequent stage of the platform, where the request's verifiability is assessed. The BC platform's second task involves cryptographically signing all securely classified requests through a series of blocks. This marks the final action carried out by the system exclusively for transactions deemed secure. When the platform generates an authentic and secure response message, it proceeds to cryptographically sign the request for enhanced security.

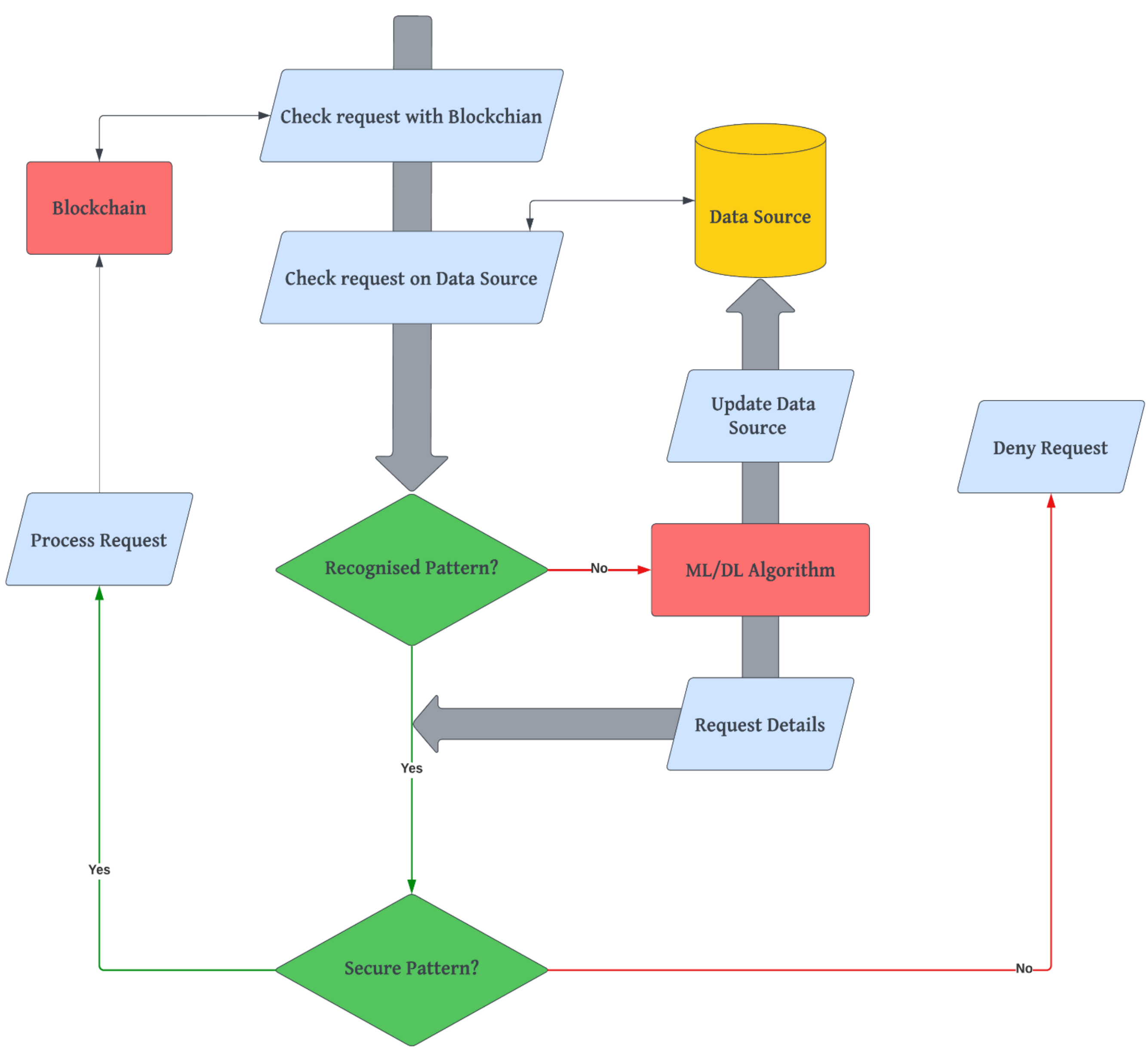

**Fig. 3** Flowchart of Blockchain-Enabled AI Security Agent.

### 3.3 Phase 2. Request Pattern Recognition Check

In its second operation, the agent checks with its data source to verify the recognition of the incoming request's pattern. A response message is then received, conveying whether the pattern is acknowledged or not. If recognized, specific security level details are provided. In case of non-recognition, the agent is informed about the lack of recognition and is supplied with details on patterns resembling the incoming request. The subsequent step involves processing the request through the DL algorithm. The data source not only verifies the pattern but also furnishes the agent with security-level details for incoming requests. With this information, acceptance or rejection of the request determined by the DL algorithm is determined by the agent. The data source consists of several entities, each of which has specific functions (Table 4).

**Table 4.** Entities of data source.

| Entity | Responsibilities |
|---|---|
| First | Save ML/DL algorithms for predicting attacks. |
| Second | Maintain details of past attacks and their patterns. |
| Third | Handle requests and match them with patterns stored in the DS. |

## 3.4 Phase 3. ML/DL Algorithm

In this phase, before training the model, we select features using the modified Whale Optimization Algorithm (WOA). This is done to reduce training time, prevent over-fitting, and improve the accuracy of the model. The key concepts of WOA are given below:

**Encircling Prey**: WOA is based on the idea of simulating the encircling behavior of whales when hunting for prey. Whales collaborate to surround their target, and this cooperative behavior is translated into optimization search strategies.

**Exploration and Exploitation**: The algorithm balances exploration and exploitation by incorporating a mechanism for both global and local search. It employs exploration in the early stages and shifts towards exploitation as the search progresses.

**Mathematical Modeling**: WOA employs mathematical equations to model the movement of whales and the updating of potential solutions. These equations guide the optimization process by simulating the hunting and encircling behaviors.

**Updating Position of Whales**: The position of each whale in the search space is updated iteratively (1). The new position (solution) is determined based on the current position, a random coefficient, and the distance to another whale (prey).

$$P_{new} = P_{rand} - A \cdot \partial \quad (1)$$

Where:

1. $P_{new}$ is the new position of the whale.
2. $P_{rand}$ is a randomly selected whale's position.
3. A is a random coefficient in the range [0, 2].
4. $\partial$ is the distance between the selected and the current whale.

**Updating Encircling Behavior**: The encircling behavior is modeled by adjusting the position based on the distance to the prey and a parameter C that controls the spiral movement (2).

$$\partial = | \delta \cdot P_{rand} - P_{prey} | \quad (2)$$

Where:

1. $P_{prey}$ is the position of the prey (best solution).

2. $\delta$ is a coefficient controlling the spiral motion.

**Global Exploration**: During the exploration phase, a random exploration/exploitation coefficient $\alpha$ is used to update the position globally (3).

$$P_{global} = P_{rand} - \alpha \cdot A \tag{3}$$

Where:

1. $P_{global}$ is the globally updated position.
2. $\alpha$ is a random exploration/exploitation coefficient.

The fundamental WOA involves whales navigating through a continuous search space, modifying their positions freely. However, when dealing with Feature Selection issues, solutions are constrained to binary values $\{0,1\}$. To address Feature Selection problems, a new variant called binary Whale Optimization Algorithm is introduced. The optimal feature combination is determined by maximizing classification accuracy while minimizing the number of selected features. Equation 4 represents the fitness function employed in the two binary versions introduced, evaluating the positions of individual whales.

$$Fitness = \lambda \Lambda_S(D) + \gamma \frac{|C - S|}{|C|} \tag{4}$$

Where:

1. $\lambda$ and $\gamma$ represent two parameters that mirror the length of the subset and the accuracy of classification.
2. C represents the number of features.
3. $S$ signifies the length of the selected feature subset.
4. $\Lambda_S(D)$ denotes the classification accuracy of the condition attribute set $S$ to the decision D.

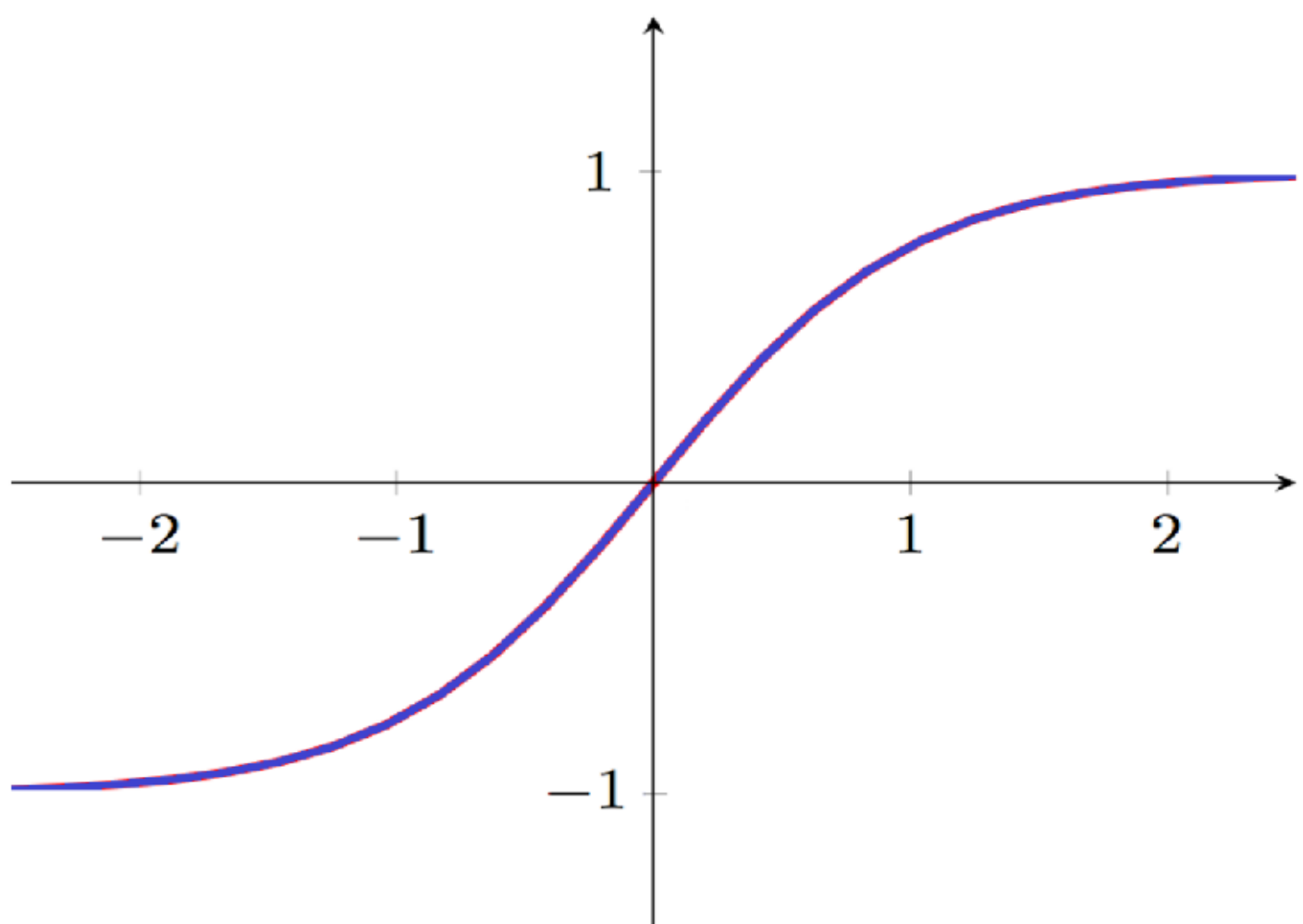

**Fig. 4** Tanh function.

Hence, (4) which focuses on the classification error rate and the selected features becomes a minimization problem, diverging from the classification accuracy and the size of unselected features. This results in the minimization problem outlined in (5).

$$Fitness = \lambda E_R(D) + \beta \frac{|R|}{|C|} \tag{5}$$

Where:

1. $E_R(D)$ is the classification error.

The proposed algorithm incorporates the tanh function (Figure 4), resembling V-shaped functions (Figure 5) as described in (6) and (7).

$$y^k = \tanh x^k \tag{6}$$

$$X_i^d = \begin{cases} 0 & \text{if rand} < S(x_i^k(t+1)) \\ 1 & \text{otherwise} \end{cases} \tag{7}$$

This adaptation enables the binary WOA algorithms to dynamically explore the feature space, ultimately aiming for the discovery of the optimal feature combination.

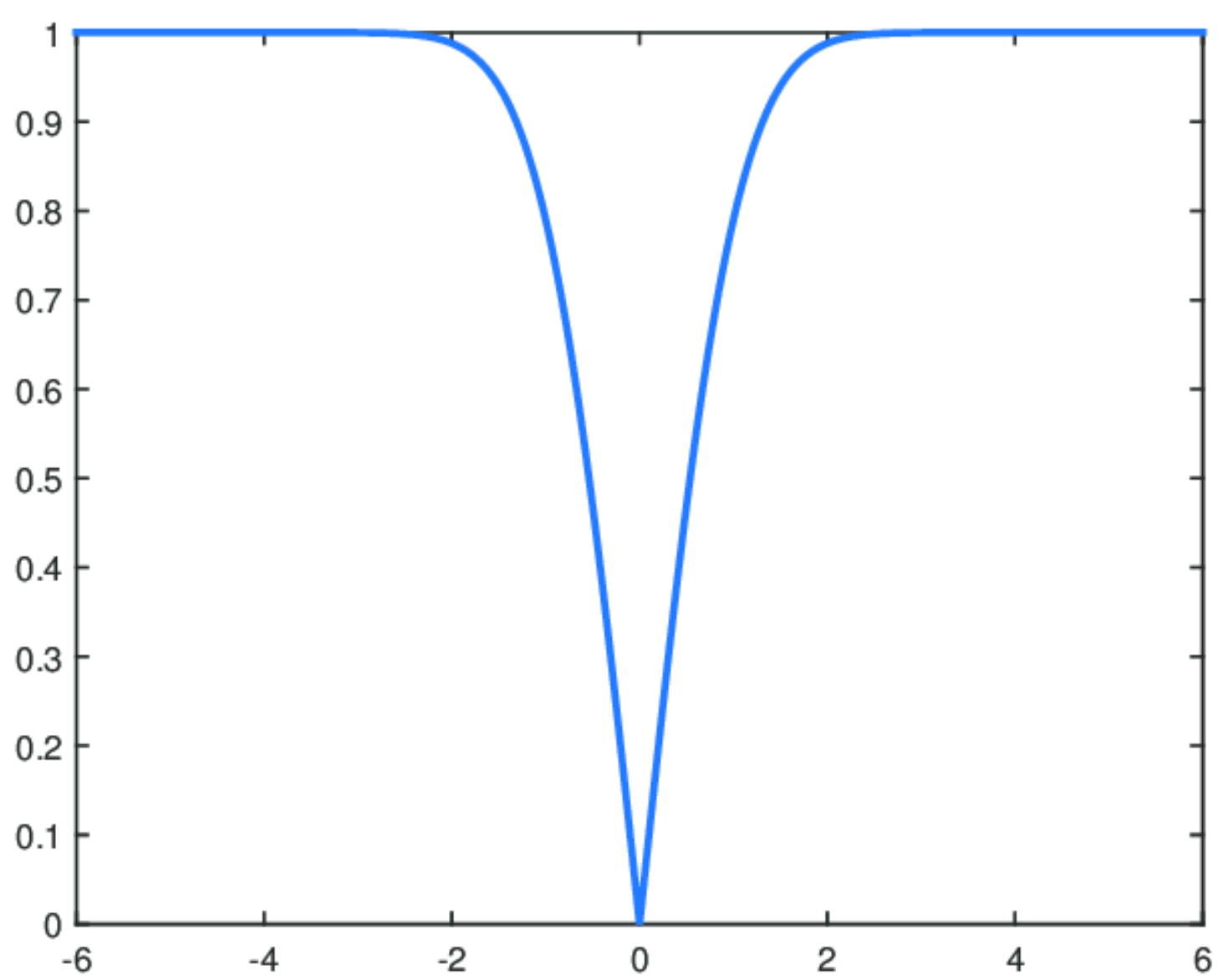

**Fig. 5** V-shaped transfer function.

Algorithm 2 illustrates the implementation of the WOA algorithm.

| **Algorithm 2.** Pseudocode for WOA. |
| --- |
| 1. Initialize population size N, maximum iterations T, and dimension D<br>2. Initialize population of search agents randomly<br>3. Evaluate fitness of initial search agents<br>4. **For** each iteration t from 1 to T **do**:<br>    5. Update parameter a (a = 2 - t * (2 / T)) |

6. **For** each search agent i **do**:
7. Update A = 2 * a * random() - a
8. Update C = 2 * random()
9. Update p = random()
10. **If** p < 0.5 **then**:
11. **If** |A| < 1 **then**:
12. Select best agent
13. Update position of agent i towards best agent
14. **Else**:
15. Select random agent
16. Update position of agent i towards random agent
17. **Else**:
18. Select best agent
19. Update position of agent i using spiral-shaped path around best agent
20. Evaluate fitness of updated search agents
21. Select the best agent as the best solution

BiLSTM networks are specifically built to understand the sequential relationships in data. The processing of input sequences in the BiLSTM network happens simultaneously in the forward and backward directions. This type of operation makes them suitable for tasks that include time dependencies. Figure 6 illustrates the architecture of a BiLSTM, which consists of multiple consecutive time steps.

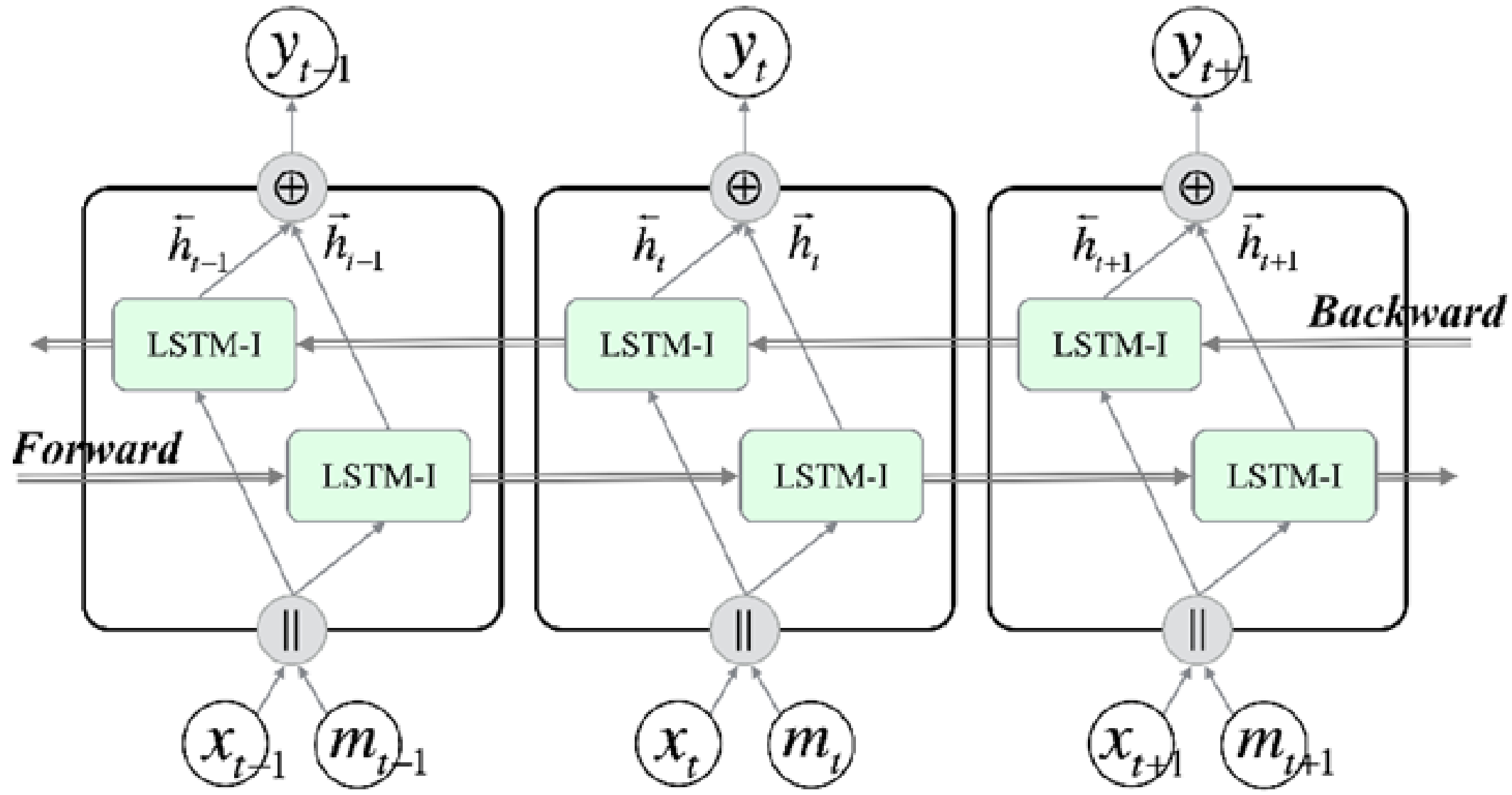

**Fig. 6** BiLSTM neural network architecture.

The network architecture consists of LSTM units that store and retrieve information over extended sequences. The BiLSTM network identifies the patterns in the data based on (8)-(10).

$$\overrightarrow{\Gamma_h}(t) = \Phi_H(\overrightarrow{W}_i X_t + \overrightarrow{\eta}_i \overrightarrow{\Gamma_h}(t-1) + \vec{b}_i) \tag{8}$$

$$\overleftarrow{\Gamma_h}(t) = \Phi_H(\overleftarrow{W}_i X_t + \overleftarrow{\eta}_i \overleftarrow{\Gamma_h}(t-1) + \overleftarrow{b}_i) \tag{9}$$

$$y(t) = \overrightarrow{Q}_i \overrightarrow{\Gamma_h}(t) + \overleftarrow{Q}_i \overleftarrow{\Gamma_h}(t) + b_y \tag{10}$$

The ultimate output vector y(t) is computed as:

$$y(t) = \omega_y(\overrightarrow{\Gamma_h} . \overleftarrow{\Gamma_h}) \tag{11}$$

The data are normalized using (12).

$$X_{normalized} = \frac{X - X_{min}}{X_{max} - X_{min}} \tag{12}$$

Where:

1. $X_{normalized}$ is the normalized value.
2. $X_{min}$ and $X_{max}$ are the minimum and maximum values of data.

The flowchart of how to train the BILSTM network is shown in Figure 7.

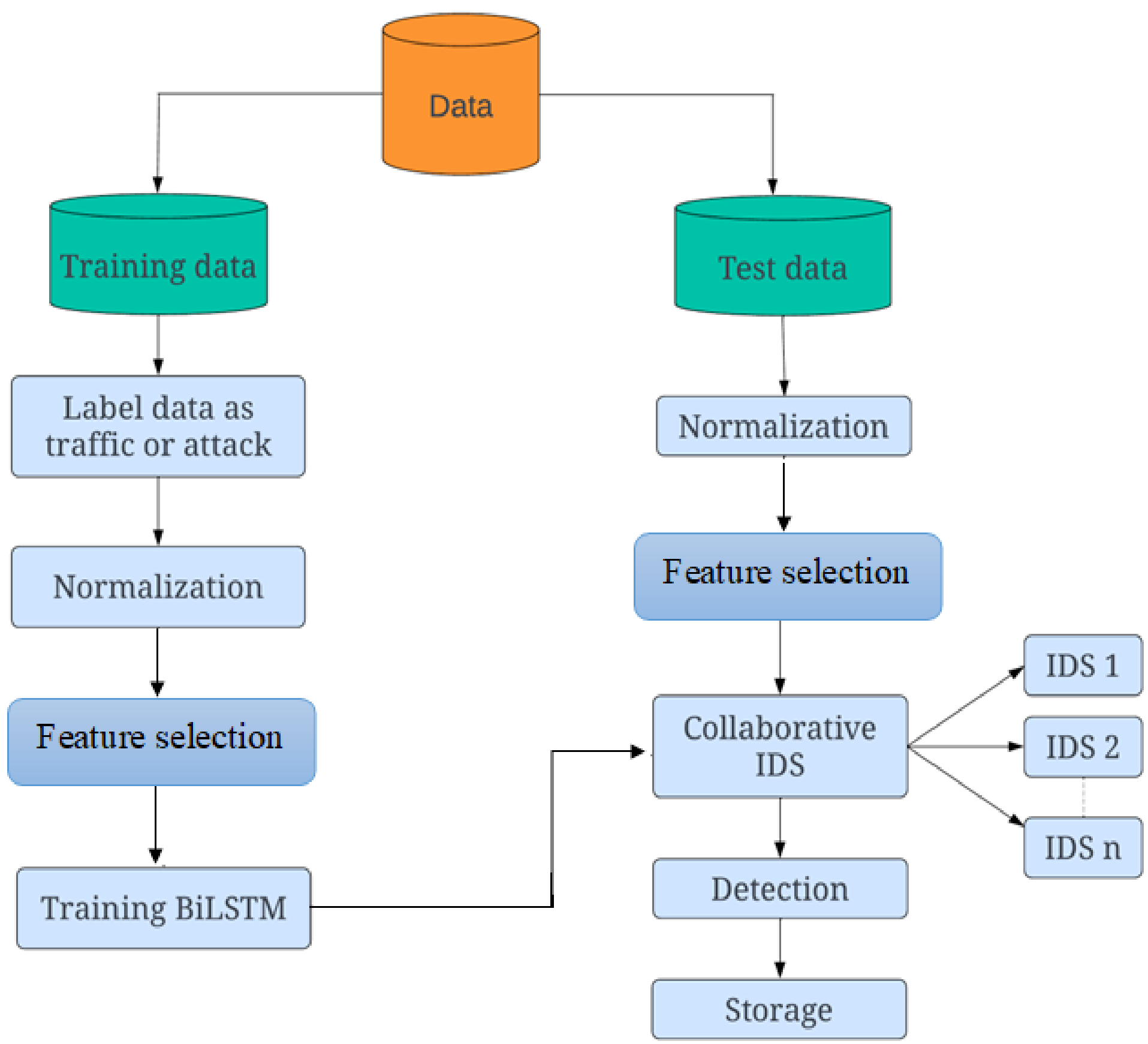

**Fig. 7** Intrusion detection-based BiLSTM.

BiLSTM networks are particularly well-suited for tasks involving sequential data due to their ability to process information in both forward and backward directions. This bidirectional nature enables BiLSTMs to capture context from both past and future states, providing a more comprehensive understanding of the data compared to unidirectional LSTMs and other sequential models. This feature is especially valuable in intrusion detection systems where the context of network activity can be crucial for accurate detection. Traditional LSTMs process data only in one direction, which can limit their ability to understand the full context of each data point in a sequence. By incorporating information from both directions, BiLSTMs can better capture dependencies and relationships within the data. This enhanced contextual understanding can lead to more accurate predictions, making BiLSTMs a superior choice for our application.

# 4 Simulation Results

The proposed method is evaluated on two distinct datasets, NSL-KDD [31], UNSW-NB15 [32], and WUSTL-EHMS-2020 [35]. Subsequently, the obtained results are compared with findings from prior works referenced as [10], [33], and [34]. This comparative analysis aims to gauge the effectiveness and performance of the proposed method compared to existing approaches on UWSN, WUSTL-EHMS-2020, and NSL-KDD datasets.

## 4.1 Experimental Setup

The experiments were conducted on a system with an Intel Core i7-10700K processor, which has 8 cores and 16 threads, operating at a base frequency of 3.80 GHz. The system was equipped with 32 GB of DDR4 RAM and a 1 TB NVMe SSD for storage. Graphics processing was handled by an NVIDIA GeForce RTX 2080 Ti with 11 GB of GDDR6 VRAM, and the operating system used was Ubuntu 20.04 LTS. For training the BiLSTM network, the following hyperparameter settings were used: a learning rate of 0.001, a batch size of 64, and the model was trained for 100 epochs using the Adam optimizer and cross-entropy loss function [21, 23, 24, 26, 28]. The activation functions applied were ReLU for the input layer, Tanh for the hidden layers, and Sigmoid for the output layer. A dropout rate of 0.5 was applied to all LSTM layers, with each layer containing 128 LSTM units, and the network comprised three layers (one input, one hidden, and one output). During the feature selection phase, the binary Whale Optimization Algorithm (WOA) was employed with a population size of 50 and a maximum of 200 iterations. The convergence criteria were set at $1 \times 10^{-6}$, and the exploration-exploitation balance parameter (α) was linearly decreased from 2 to 0 over the iterations.

## 4.2 Performance metrics

**Precision**: measures how well a model performs in correctly identifying positive cases without including false positives (13).

$$Prec. = \frac{TP}{FP + TP} \tag{13}$$

**Recall**: quantifies how well a model identifies positive cases without missing any true positives. (14).

$$Rec. = \frac{TP}{FP + FN} \tag{14}$$

**F1-Score**: provides a comprehensive evaluation of a model's performance in terms of both positive predictions and capturing all relevant positive instances (15).

$$F1-Score=\frac{2\times Prec.\times Rec.}{Prec.+Rec.} \tag{15}$$

**Accuracy**: quantifies the overall correctness of predictions, which is calculated by (16).

$$Accuracy=\frac{TP+TN}{Total\ Instances} \tag{16}$$

**Detection rate**: The metric assesses the percentage of positive cases accurately detected by a system (17).

$$DR=\frac{TP}{TP+FN} \tag{17}$$

**False alarm rate:** measures the rate at which the system produces false positive predictions or detections (18).

$$FAR=\frac{FP}{FP+TN} \tag{18}$$

## 4.3 Analysis of Results

Figure 8 illustrates the variations in training and validation accuracy, as well as training and validation loss, for two distinct datasets: NSL-KDD and UNSW-NB15. The analysis of the results is given below:

**(a) Accuracy of training and validation in the NSL-KDD dataset:** The training accuracy starts at 85% and gradually increases with each epoch, reaching 97% at epoch 40. The validation accuracy also starts at 86% and shows a similar increasing trend, reaching 99.5% at epoch 40. Both training and validation accuracies consistently improve over time, indicating that the BiLSTM network is effectively learning and generalizing patterns in the NSL-KDD dataset.

**(b) Loss of training and validation in the NSL-KDD dataset:** The loss for both training and validation starts relatively high and gradually decreases with each epoch. As the epochs progress, the loss decreases, indicating that the BiLSTM network is minimizing the error between predicted and actual values. The decreasing loss values suggest that the network is learning and improving its predictive capabilities on the NSL-KDD dataset.

**(c) Accuracy of training and validation in the UNSW-NB15 dataset:** The training accuracy starts at 83% and increases steadily, reaching 95% at epoch 40. The validation accuracy begins at 84% and shows a similar increasing trend, reaching 97.5% at epoch 40. Both training and validation accuracies demonstrate improvement, but the rate of improvement slows down after epoch 25. This indicates that the BiLSTM network is learning the patterns in the UNSW-NB15 dataset, but the rate of improvement plateaus after a certain point.

**(d) Loss of training and validation in the UNSW-NB15 dataset:** The loss for training and validation starts relatively high and decreases gradually with each epoch. The decreasing loss values indicate that the BiLSTM network effectively minimizes the error between predicted and actual values in the UNSW-NB15 dataset. However, similar to the accuracy, the rate of improvement in loss slows down after epoch 25.

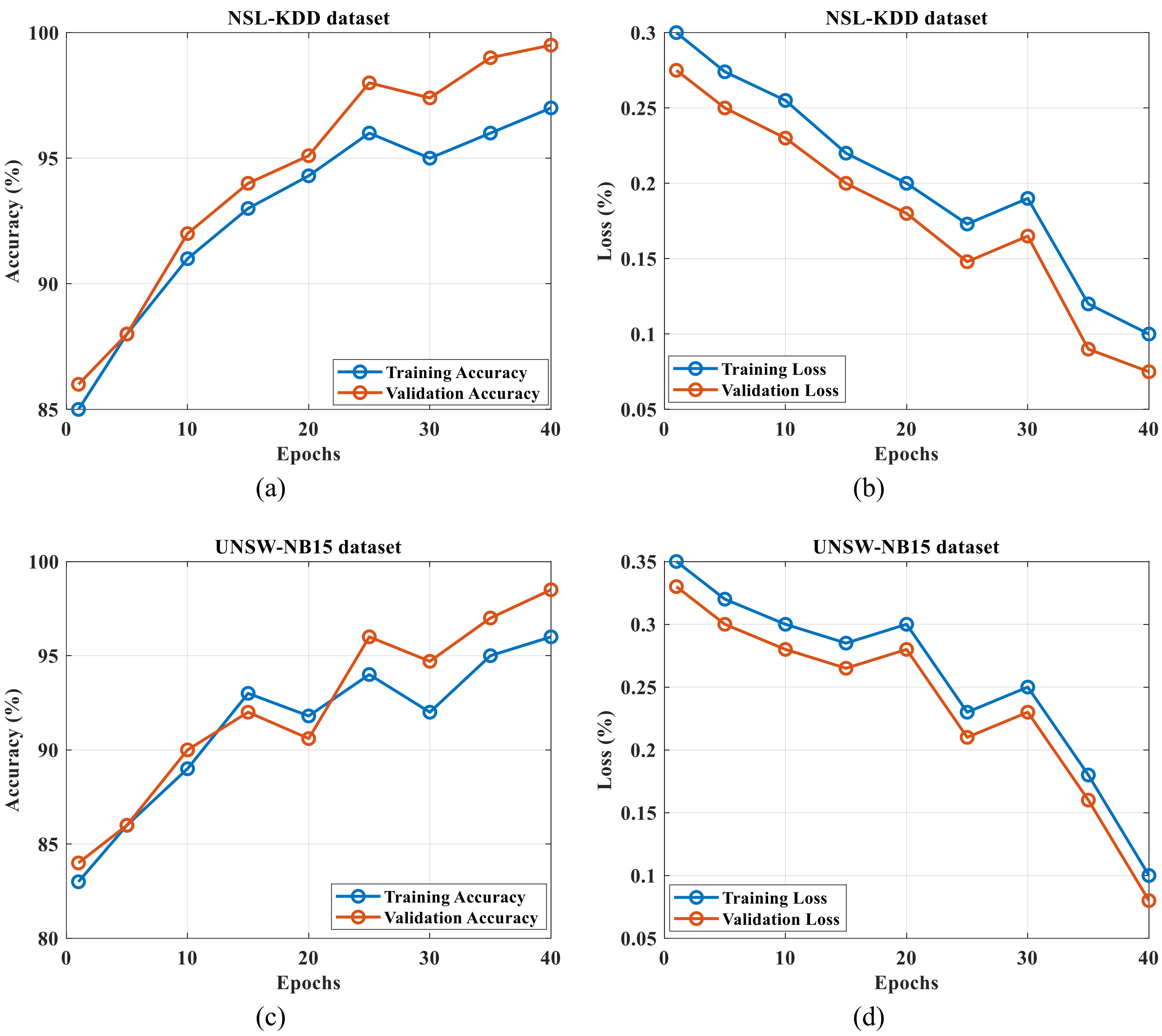

**Fig. 8** (a) Accuracy and (b) Loss of training and validation in the NSL-KDD dataset. (c). Accuracy and (d) training and validation loss in the UNSW-NB15 dataset.

Figure 9, Table 5, and Table 6 compare the proposed method and other methods in terms of precision, recall, accuracy, and F1-score on the NSL-KDD dataset. The analysis of the results is given below:

**Precision, Recall, and F1-score:** As shown in Figure 9, the proposed method and other methods have performed better on the NSL-KDD dataset than the UNSW-NB15 dataset. The reason for this is the high number of attacks in the UNSW-NB15 data set, which is more complex than the NSL-KDD data. The feature selection technique we used in this paper provided BILSTM with good generalization power. In addition, the BiLSTM network has identified attack patterns better than other methods due to the bidirectional nature of the calculations.

**Accuracy:** The feature selection technique helps select the most relevant and informative features, reducing noise and improving the model's ability to distinguish between classes. The BiLSTM network, with its ability to capture both past and future context, increases the model's understanding of temporal dependencies and improves its predictive capabilities. For this reason, the proposed model has shown better performance.

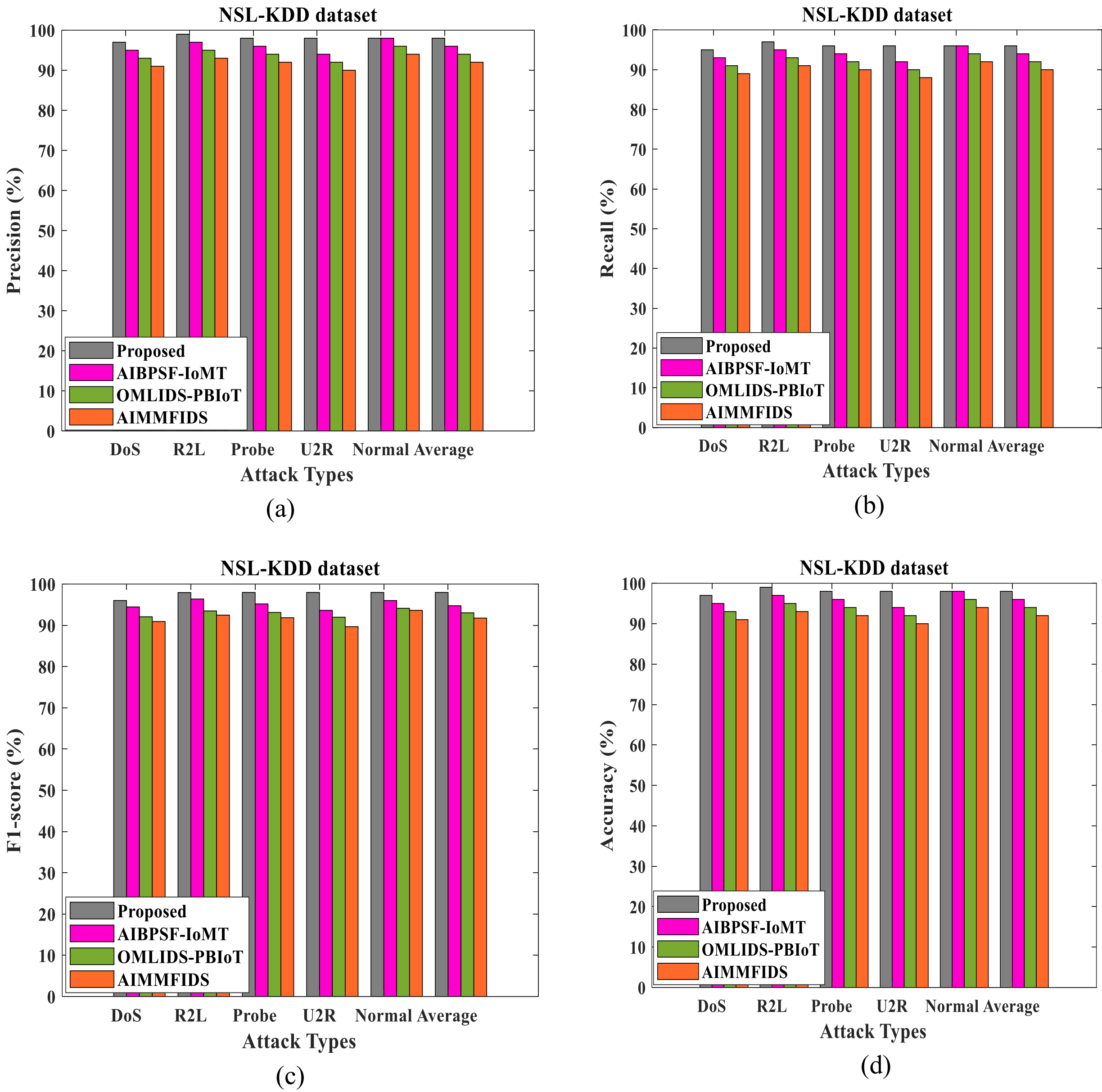

**Fig. 9** Comparison of (a) Precision, (b) Recall, (c) F1-score, and (d) accuracy of the proposed method and three other methods in the NSL-KDD dataset.

**Table 5.** Precision Scores of Various Methods Applied to the NSL-KDD Dataset.

| Attack type | **Proposed** | AIBPSF-IoMT | OMLIDS-PBIoT | AIMMFIDS |
|---|---|---|---|---|
| DoS | **97** | 95 | 93 | 91 |
| R2L | **99** | 97 | 95 | 93 |
| Probe | **98** | 98 | 94 | 92 |
| U2R | **98** | 94 | 92 | 90 |
| Normal | **98** | 96 | 96 | 94 |
| Average | **98** | 96 | 94 | 92 |

**Table 6.** Recall Scores of Various Methods Applied to the NSL-KDD Dataset.

| Attack type | **Proposed** | AIBPSF-IoMT | OMLIDS-PBIoT | AIMMFIDS |
|---|---|---|---|---|
| DoS | **95** | 93 | 91 | 89 |
| R2L | **97** | 95 | 93 | 91 |
| Probe | **96** | 94 | 92 | 90 |
| U2R | **96** | 92 | 90 | 88 |
| Normal | **96** | 96 | 94 | 92 |
| Average | **96** | 94 | 92 | 90 |

Figure 10, Table 7, and Table 8 compare the proposed method and other methods in terms of precision, recall, accuracy, and F1-score on the UNSW-NB15 dataset. The analysis of the results is given below:

In evaluating intrusion detection methods on the UNSW-NB15 dataset, the proposed approach, which integrates feature selection techniques and a BILSTM network, consistently outperforms alternative methods. This is evident through examining precision, recall, accuracy, and F1-score metrics across various attack types. The proposed method exhibits remarkable precision, indicating high correctness in predicting attacks and elevated recall values, showcasing its proficiency in capturing a significant proportion of actual attacks. Comparatively, other methods, such as AIBPSF-IoMT, OMLIDS-PBIoT, and AIMMFIDS, though competitive, demonstrate slightly lower precision, recall, and accuracy values. The observed superiority of the proposed method can be attributed to several factors.

The effective feature selection techniques enhance the model's ability to discern relevant information amidst noise. Additionally, the BILSTM network's capacity to capture temporal dependencies in sequential data proves advantageous in the context of intrusion detection. Upon comparison with the NLS-KDD dataset, it becomes apparent that the precision, recall, and accuracy values in the UNSW-NB15 dataset are generally smaller. This discrepancy may be attributed to the increased complexity of the UNSW-NB15 dataset, characterized by diverse attack patterns and potential class imbalances. The proposed method, adept at handling these intricacies, demonstrates consistent and superior performance across various attack scenarios.

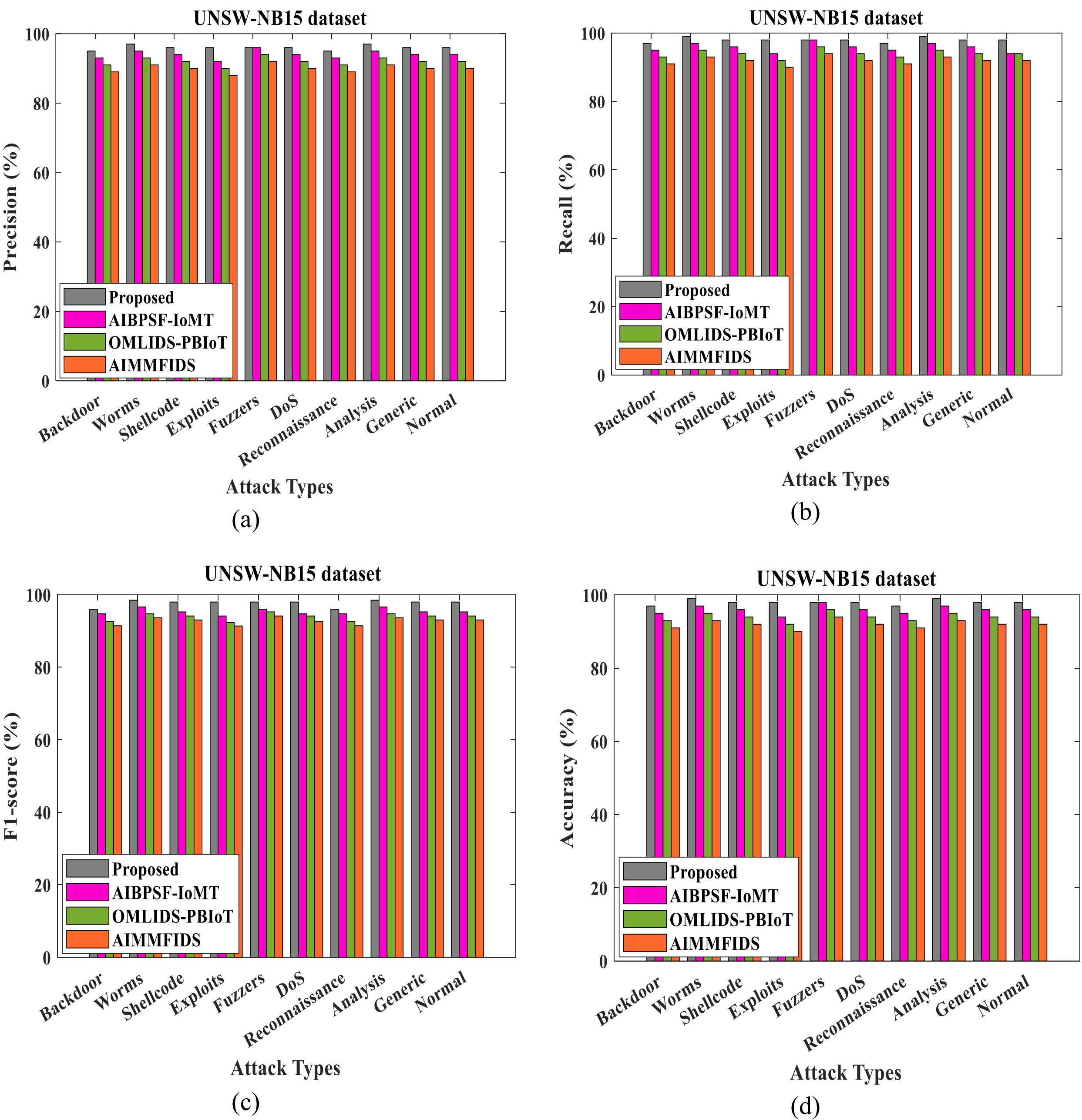

**Fig. 10** Comparison of (a) Precision, (b) Recall, (c) F1-score, and (d) accuracy of the proposed method and three other methods in the UNSW-NB15 dataset.

**Table 7.** Precision Scores of Various Methods Applied to the UNSW-NB15 Dataset.

| Attack type | **Proposed** | AIBPSF-IoMT | OMLIDS-PBIoT | AIMMFIDS |
|---|---|---|---|---|
| Backdoor | **96** | 94 | 92 | 90 |
| Worms | **98** | 96 | 94 | 92 |
| Shellcode | **97** | 95 | 93 | 93 |
| Exploits | **97** | 93 | 91 | 89 |
| Fuzzers | **97** | 97 | 95 | 91 |
| DoS | **97** | 95 | 93 | 91 |
| Reconnaissance | **96** | 94 | 92 | 90 |
| Analysis | **98** | 96 | 94 | 92 |
| Generic | **97** | 95 | 93 | 91 |
| Normal | **97** | 95 | 93 | 91 |
| Average | **97** | 95 | 93 | 91 |

**Table 8.** Recall Scores of Various Methods Applied to the UNSW-NB15 Dataset.

| Attack type | **Proposed** | AIBPSF-IoMT | OMLIDS-PBIoT | AIMMFIDS |
|---|---|---|---|---|
| Backdoor | **94** | 92 | 90 | 88 |
| Worms | **96** | 94 | 92 | 90 |
| Shellcode | **95** | 93 | 91 | 89 |
| Exploits | **95** | 91 | 89 | 87 |
| Fuzzers | **95** | 95 | 93 | 91 |
| DoS | **95** | 93 | 91 | 89 |
| Reconnaissance | **94** | 92 | 90 | 88 |
| Analysis | **96** | 94 | 92 | 90 |
| Generic | **95** | 93 | 91 | 89 |
| Normal | **95** | 93 | 91 | 89 |
| Average | **95** | 93 | 91 | 89 |

Figure 11, Table 9, and Table 10 show the performance of different intrusion detection methods in terms of detection rate and false alarm rate in different attack percentages. The proposed method consistently demonstrates a high detection rate, showcasing its effectiveness in identifying intrusions even as the percentage of attacks increases. Despite a rise in the false alarm rate at higher attack percentages, the method maintains a commendable balance between accurate detections and false positives. AIBPSF-IoMT exhibits a reasonable detection rate, though slightly lower than the proposed method, and a moderate false alarm rate. Similar to the proposed method, the detection rate decreases with an increase in attack percentages, and the false alarm rate sees a noticeable uptick at higher attack percentages, indicating some instances of false positives. OMLIDS-PBIoT maintains a moderate to high detection rate, showing resilience to varying attack percentages. However, like AIBPSF-IoMT, the false alarm rate increases gradually as the attack percentage rises. AIMMFIDS, while effective in detection, experiences a more pronounced decrease in the detection rate as the attack percentage increases. Furthermore, the false alarm rate for AIMMFIDS rises significantly at higher attack percentages, indicating a higher likelihood of false positives.

**Table 9.** Detection Rates of Various Methods Applied to the NSL-KDD Dataset.

| Attack type | AP=30 | AP=40 | AP=50 | AP=60 | AP=70 | AP=80 |
|---|---|---|---|---|---|---|
| **Proposed** | **0.98** | **0.97** | **0.95** | **0.92** | **0.89** | **0.87** |

| AIBPSF-IoMT | 0.96 | 0.94 | 0.92 | 0.90 | 0.85 | 0.82 |
|---|---|---|---|---|---|---|
| OMLIDS-PBIoT | 0.96 | 0.93 | 0.90 | 0.86 | 0.83 | 0.81 |
| AIMMFIDS | 0.94 | 0.93 | 0.89 | 0.85 | 0.82 | 0.79 |

**Table 10.** Detection Rates of Various Methods Applied to the UNSW-NB15 Dataset.

| Attack type | AP=30 | AP=40 | AP=50 | AP=60 | AP=70 | AP=80 |
|---|---|---|---|---|---|---|
| **Proposed** | **0.98** | **0.95** | **0.93** | **0.91** | **0.88** | **0.85** |
| AIBPSF-IoMT | 0.95 | 0.93 | 0.90 | 088 | 0.84 | 0.81 |
| OMLIDS-PBIoT | 0.95 | 0.92 | 0.89 | 0.85 | 0.83 | 0.81 |
| AIMMFIDS | 0.93 | 0.93 | 0.88 | 0.85 | 0.80 | 0.78 |

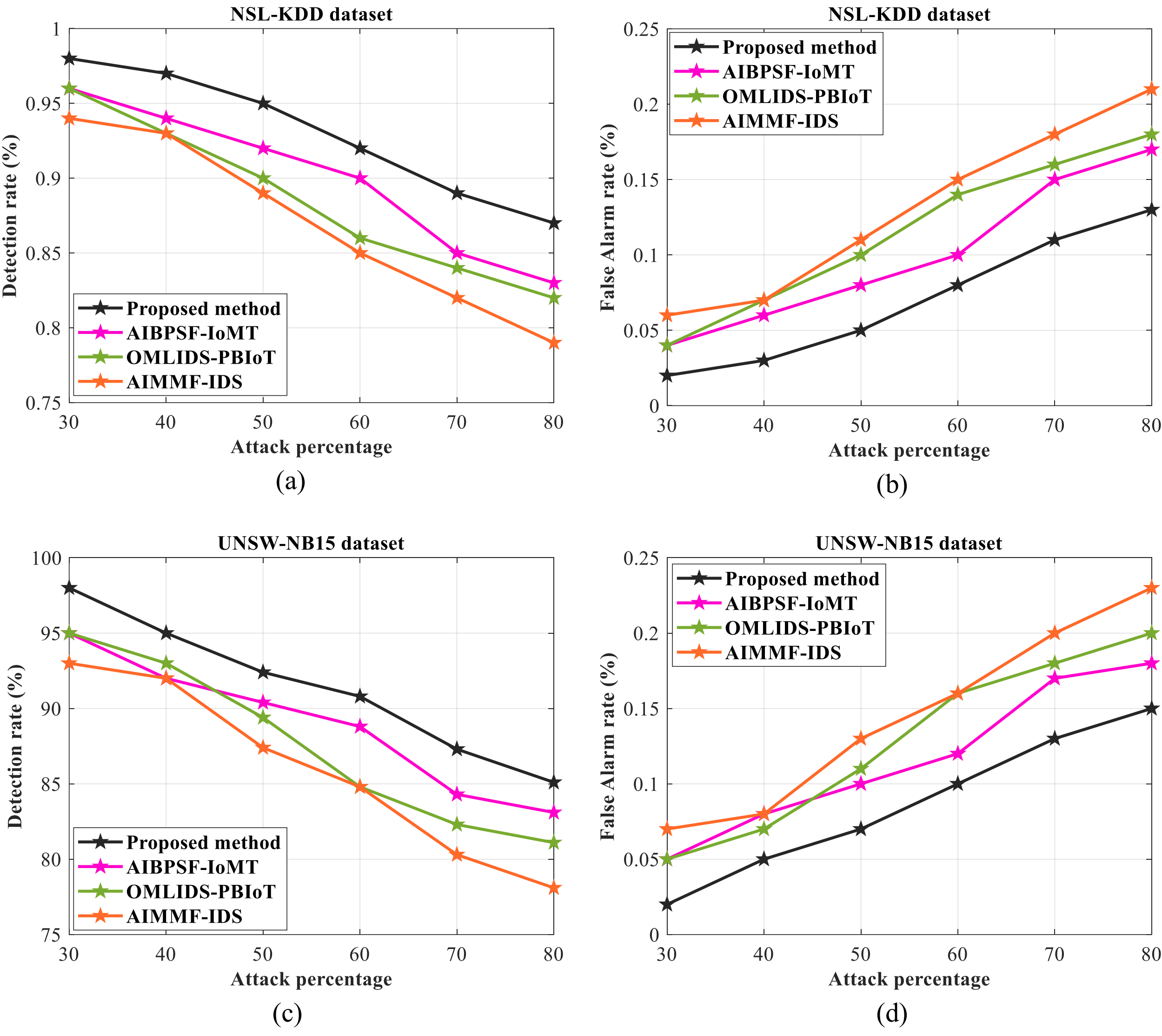

**Fig. 11** (a) Detection rate at different attack percentages in the NSL-KDD dataset. (b) False Alarm rate at different attack percentages in the NSL-KDD dataset. (c) Detection rate at different attack percentages in the UNSW-NB15 dataset. (d) False Alarm rate at different attack percentages in the UNSW-NB15 dataset.

Table 11 presents the comparative analysis of the proposed method and three existing methods: AIBPSF-IoMT, OMLIDS-PBIoT and AIMMFIDS. Evaluation criteria include accuracy, precision, recall, F1 score, detection rate, and false alarm rate. This table clearly shows the superior performance of the proposed method in all criteria, especially in achieving high detection rate and low false alarm rate. which are critical for effective intrusion detection in IoT environments. This evaluation was performed on the WUSTL-EHMS-2020 dataset provided for medical work.

**Table 11.** Comparative analysis of the proposed method and existing methods in WUSTL-EHMS-2020 dataset.

| Method | Accuracy (%) | Precision (%) | Recall (%) | F1-Score (%) | Detection Rate (%) | False Alarm Rate (%) |
|---|---|---|---|---|---|---|
| **Proposed Method** | **98.2** | **97.8** | **98.5** | **98.1** | **97.6** | **1.2** |
| AIBPSF-IoMT | 95.5 | 94.2 | 95.8 | 95.0 | 94.7 | 2.5 |
| OMLIDS-PBIoT | 96.1 | 95.0 | 96.3 | 95.6 | 95.2 | 2.0 |
| AIMMFIDS | 94.8 | 93.5 | 95.0 | 94.2 | 93.8 | 3.0 |

## 4.4 Statistical Significance Testing

While our comparative analysis with recent methods is thorough, we have enhanced it by including statistical significance tests and using cross-validation to ensure the robustness of our results. To provide a reliable evaluation of our proposed method, we employed k-fold cross-validation. This approach involves dividing the dataset into k equally sized folds, training the model on k-1 folds, and testing it on the remaining fold. This process is repeated k times, with each fold being the test set once. The performance metrics are averaged across all k trials for a more stable and generalizable estimate. To verify that the observed performance improvements are statistically significant and not due to random variation, we conducted the following statistical tests:

1. **Paired t-test**: This test compares the mean performance of our proposed method with each baseline method across the cross-validation folds, evaluating whether the mean difference is significantly different from zero.
2. **Wilcoxon signed-rank test**: As a non-parametric alternative, this test compares the ranks of performance differences between our method and the baseline methods. It is beneficial for non-normally distributed data or small sample sizes.

We evaluated the methods using key performance metrics such as accuracy, precision, recall, F1-score, and AUC-ROC. We performed k-fold cross-validation (with k=10) to obtain robust performance estimates for our proposed and baseline methods. We conducted paired t-tests to compare the mean performance of our method against each baseline method across the cross-validation folds. We conducted Wilcoxon signed-rank tests to validate the results from the paired t-tests and account for non-normal distributions. A confidence level of 95% (p-value < 0.05) was used to determine statistical significance. The results of the statistical significance tests are summarized in Table 12.

**Table 12.** Results of Statistical Significance Tests for Performance Comparison.

| Metric | Paired t-test p-value | Wilcoxon p-value | Significance ($p < 0.05$) |
|---|---|---|---|
| Accuracy | 0.001 | 0.002 | YES |
| Precision | 0.003 | 0.004 | YES |
| Recall | 0.020 | 0.025 | YES |
| F1-score | 0.005 | 0.006 | YES |
| AUC-ROC | 0.000 | 0.001 | YES |

## 5 Conclusion and Future Work

This paper introduced an innovative approach to address the critical challenge of medical data security within the IoT landscape. Our method seamlessly integrates BC technology and advanced machine learning techniques, providing a robust framework to fortify the confidentiality and integrity of sensitive healthcare information. Through a meticulous three-phase process, incorporating Blockchain-Enabled Request and Transaction Encryption, Request Pattern Recognition Check, and Feature Selection coupled with the BiLSTM network, we aimed to enhance the security posture of IoT-based healthcare systems. We compared the performance of the proposed method with three methods: AIBPSF-IoMT, OMLIDS-PBIoT, and AIMMFIDS. We achieved significant improvement in various criteria: Accuracy (2%), Precision (2%), Recall (2%), Attack Detection rate (5%) and False Alarm rate (3%). The proposed method is designed to be computationally efficient and scalable, ensuring its practical applicability in real-world healthcare settings. By leveraging advanced techniques such as the modified WOA for feature selection and the BiLSTM network for intrusion detection, our approach minimizes computational overhead while maintaining high accuracy. This efficiency enables the method to be implemented on existing hardware within healthcare institutions without necessitating extensive upgrades. Furthermore, our method is highly scalable and capable of handling increasing volumes of healthcare data through blockchain technology for secure, decentralized data management. The proposed method can seamlessly integrate with existing electronic health record (EHR) systems and medical devices, ensuring interoperability through adherence to established standards such as HL7 and FHIR. This compatibility facilitates easy adoption and maximizes our method's impact on enhancing healthcare data security and privacy. While our method is computationally efficient, further optimization can be explored. Future work could investigate using parallel computing techniques and hardware accelerations such as GPUs and TPUs to enhance processing speed and reduce computational costs. As the volume of healthcare data continues to grow, scalability remains a critical focus. Future research could explore advanced blockchain technologies, such as sharding or layer two solutions, to efficiently manage large datasets. Additionally, leveraging cloud computing resources for distributed data storage and processing can further improve scalability.

## Declarations

### Funding

No funds

### Conflicts of interest

Conflict of Interest is not applicable in this work.

### Availability of data and material

Not applicable

### Ethical approval

Not applicable

### CRediT author statement

**Behnam Rezaei Bezanjani:** Conceptualization, Data curation, Investigation, Methodology, Resources, Software, Validation, Visualization, Writing - original draft, Writing - review & editing.

**Seyyed Hamid Ghafouri:** Formal analysis, Investigation, Methodology, Resources, Software, Supervision, Visualization, Writing - review & editing.